\documentclass[epj]{svjour}
\usepackage{graphics}
\begin{document}
\title{On temperature dependence of quarkonium correlators}
\author{P\'eter Petreczky\inst{1} 
}                     
\offprints{}          
\institute{Physics Department, Brookhaven National Laboratory, 
Upton NY 11973 USA}
\date{Received: date / Revised version: date}
%
\abstract{
I discuss the temperature dependence of quarkonium correlators
calculated in lattice QCD. I show that the dominant source of
the temperature dependence comes from the zero mode contribution, while
the the temperature dependence associated with melting of bound states
is quite small.
I study the zero mode contribution quantitatively for
different quark masses and show that it is well described by a 
quasi-particle model with temperature dependent heavy quark mass.
As a byproduct an estimate of the medium dependence of the heavy quark mass
is obtained.
\PACS{
      {12.38.Mh}{Quark-gluon plasma }   \and
      {12.38.Gc}{Lattice QCD calculations}
     } 
} 
\maketitle
\section{Introduction}
\label{intro}
Quarkonium properties at finite temperature received
considerable interest since the famous conjecture by Matsui
and Satz that color screening will lead to quarknonium 
dissociation, which in turn can signal the onset of deconfinement
in heavy ion collisions \cite{MS86}. This problem has been studied
using potential models with screened temperature dependent potential
\cite{karsch88,ropke88,hashimoto88,digal01a,digal01b,digal02,shuryak04,wong04,blaschke,alberico,mocsy06,rapp,alberico06,mocsy07a,mocsy07b,mocsy_sqm07,mocsy_zj75}. 
The early studies used different phenomenological parametrization, 
while more recent ones use the so-called singlet free energy of a
static $Q\bar Q$ pairs as an input. Since the lattice studies of the
singlet free energy show quite strong screening 
(see  Refs. \cite{me_hard04,olaf_cpod07}) potential models typically
predict large in-medium modification of quarkonium properties and/or
its dissolution. 

In-medium quarkonium properties 
are encoded in the corresponding spectral functions which in turn
are related to 
meson (current-current) correlation functions in Euclidean time.
Euclidean correlation functions can be calculated on the lattice.
Therefore one can hope to learn something about in-medium quarkonium
properties by studying the temperature dependence of meson correlators
on the lattice. In particular, attempts to reconstruct     
quarkonium spectral functions
using the Maximum Entropy \\
Method (MEM) were presented in Refs. 
\cite{umeda02,asakawa04,datta04,doi,me_sqm06,jako07,swan}. While at zero temperature it is
possible to reconstruct the basic features of quarkonium spectral
functions the application of MEM above the deconfinement temperature
turns out to be problematic \cite{jako07}

The study of the spectral functions using MEM indicated that 1S charmonia
may survive in the deconfined phase till temperatures as high as
$(1.6-2.2)T_c$  \cite{asakawa04,datta04}. This result is closely
related to the small temperature dependence of the Euclidean correlator in
the pseudo-scalar channel \cite{datta04,jako07}. On the other hand
the large temperature dependence of the scalar and axial-vector
correlators has been interpreted as evidence for the expected 
dissolution of the P-state charmonium. However, it has been pointed
out that zero mode contribution, i.e. the contribution from
the spectral functions at very small frequencies 
could lead to  large temperature
dependence of the Euclidean correlators \cite{derek05,umeda07}.
Therefore in order to understand quarkonium properties
at finite temperature and their relation to collective effects, e.g.
screening it is important to get a detailed understanding of the 
temperature dependence of the Euclidean correlators, including the
role of the zero mode contribution. In this paper I present a 
study of the temperature dependence of quarkonium correlators 
calculated in quenched approximation on isotropic lattices.
Preliminary results of this study have been reported in conference proceedings
\cite{panic05,me_sqm07,me_qm08}. The rest of the paper is organized as follows.
I section 2 I discuss some general features of the Euclidean correlators and
spectral functions. Section 3 contains some details of the numerical calculations
as well as the discussion of the temperature dependence of the Euclidean correlators
in different channels. In section 4 the extraction of the zero mode contribution is
discussed and its temperature dependence is compared with a simple quasi-particle
model. Finally section 5 contains the conclusions. 

\section{Quarkonium spectral functions and Euclidean correlators}
\label{sec:1}
Let us consider correlation functions of meson currents $\bar q \Gamma q$,
where $\Gamma=1,~\gamma_5, \gamma_{\mu}, \gamma_5 \gamma_{\mu}$, i.e.
scalar, pseudo-scalar, vector and axial-vector currents. The relation of
these currents to different meson state is shown in Table \ref{tab1}.
Because correlation functions in Euclidean time can be considered as
analytic continuation of the real time correlators one has the 
following relation between the Euclidean correlators and the spectral
functions (e.g. see the discussion in Refs. \cite{datta04,me_sqm06,jako07} )
\begin{equation}
 G^i(\tau, T) = \int_0^{\infty} d \omega
\sigma^i(\omega,T)  \frac{\cosh(\omega(\tau-1/2 T))}{\sinh(\omega/2 T)}.
\label{spectral}
\end{equation}
\begin{table}
\caption{Quarkonium states and coefficients for the generalized susceptibilities in Eq. (\ref{gen_susc})
in different channels.}
\begin{tabular}{lllllll}
\hline
$\Gamma$ & $^{2S+1}L_{J}$ & $J^{PC}$ & particle & $a^i$ & $b^i$ & $c^i$\\
\hline
$\gamma_{5}$ & $^{1}S_{0}$ & $0^{-+}$ & $\eta_c$, $\eta_b$   & 0 & 0 & 0\\
$\gamma_{i}$ & $^{3}S_{1}$ & $1^{--}$ & $J/\psi$, $\Upsilon$ & 0 & 0 & 1\\
$\gamma_0$  & - & - & - & -1 & 0 & 0 \\
$1$ & $^{3}P_{0}$ & $0^{++}$ & $\chi_{c0}$, $\chi_{b0}$      & 0 & 1 & 0\\
$\gamma_{5}\gamma_{i}$ & $^{3}P_{1}$ & $1^{++}$ & $\chi_{c1}, \chi_{b1}$ & 1 & 2 & 0\\
\hline
\end{tabular}
\label{tab1}
\end{table} 
\begin{table*}
\caption{Parameters used in the numerical calculations and the renormalization 
constants for different currents. 
The last column gives the constituent quark mass estimated in potential model
(see the main text).
The pseudo-scalar renormalization constants
$Z_{ps}$ were estimated at the scale equal to the constituent quark mass. 
The scalar renormalization constants $Z_{sc}$ were estimated at scales equal to
half the constituent quark mass, constituent quark mass and twice the constituent quark mass (from left to right).}
\label{tab2}    
\begin{tabular}{lllllllllllll}
\hline\noalign{\smallskip}
$\beta$ & $a$ [fm] &  $\kappa$ & $a m_0$ & Vol. &\#conf. & $Z_{ax}$ & $Z_{vc}$ & $Z_{ps}$ &  $Z_{sc}$               & $M_{\eta}$ [GeV] & $m$ [GeV]\\
\noalign{\smallskip}\hline\noalign{\smallskip}
 6.499  & .0450   & 0.1300   &  0.1583 & $48^3\times 24$ & 50  & 1.003 & 0.975 & 0.681 & 0.669, 0.745, 0.822  & 2.622(50)     & 0.95 \\
        &         & 0.1279   &  0.2214 &                 &     & 1.070 & 1.040 & 0.766 & 0.757, 0.838, 0.920  & 3.271(50)     & 1.34 \\
        &         & 0.1234   &  0.3640 &                 &     & 1.221 & 1.188 & 0.928 & 0.924, 1.018, 1.112  & 4.495(10)     & 2.00 \\
\hline 
 6.640  & .0374   & 0.1290   &  0.1821 & $48^3\times 24$ & 100 & 1.034 & 1.007 & 0.808 & 0.726, 0.801, 0.876  & 3.332(270)    & 1.34 \\ 
\hline 
 7.192  & .0189   & 0.13114  &  0.0916 & $48^3\times 64$ & 100 & 0.957 & 0.936 & 0.681 & 0.670, 0.727, 0.785  & 3.619(52)     & 1.52 \\
        &         & 0.1209   &  0.4146 & $40^3\times 40$ & 80  & 1.296 & 1.283 & 1.060 & 1.060, 1.139, 1.218  & 10.632(81)    & 5.21 \\
\noalign{\smallskip}\hline
\end{tabular}
\end{table*}

Here index $i$ labels different channels, $i=sc, ps, vc$ and $ax$
for scalar, pseudo-scalar, vector and axial-vector respectively.
In the vector and axial-vector channels we sum over the three
spatial components.
Quarkonium spectral functions at zero temperature contain information
about all states with given quantum number, which contain a heavy quark anti-quark pair 
(both bound states and scattering states). At finite temperature, however, they also
describe a scattering of the external probe (e.g. a virtual photon) off a heavy
quark from the medium \cite{derek05,aarts05}. This is an analog of the Landau damping
and gives a contribution to the spectral function below the light cone ($\omega<k$).
In the limit of zero external momentum, $k \rightarrow 0$ it
becomes $\chi^i(T)\omega \delta(\omega)$
in the free theory. The generalized susceptibilities $\chi^i(T)$ 
were calculated in Ref. \cite{aarts05}
in the free theory for all channels:
\begin{equation}
\chi^i(T)=\frac{6}{\pi} \int_0^{\infty} \left( a^i+b^i \frac{m^2}{E_p^2}+c^i \frac{p^2}{E_p^2}\right)
\left( -\frac{\partial n_F(E_p)}{\partial E_p}\right).
\label{gen_susc}
\end{equation}
Here $E_p=p^2+m^2$ and $n_F(E_p)=(\exp(E_p/T)+1)^{-1}$. The coefficients $a^i$, $b^i$ and $c^i$
are given in Table \ref{tab1} for different channels. For the temporal component of the vector
correlator the generalized susceptibility is just the usual quark number susceptibility (up to
the minus sign).
Interaction of the incoming (outgoing) quark or anti-quark 
with the medium ( multiple re-scattering)
will lead to broadening of the delta function which becomes a Lorentzian \cite{derek05}:
\begin{equation}
\delta(\omega) \rightarrow \eta \frac{1}{\pi}\frac{1}{\omega^2+\eta^2}.
\end{equation}
The width of the Lorentzian $\eta \sim D^{-1} T/m$
is always small for heavy quarks $m \gg T$ because 
the diffusion constant $D$ which is proportional to the mean-free path  
cannot be much smaller than the inverse temperature $1/T$ \cite{derek05}. 
The area under the peak at $\omega \simeq 0$ is $T \chi^i(T)$ and does not
depend on $D$ \cite{derek05}.
Because the quark anti-quark pair 
contribute to the spectral functions only at energies $\omega > 2m$ quarkonium spectral functions
can be written in the form 
\begin{equation}
\sigma^i(\omega,T)=\sigma^i_{\rm high}(\omega,T)+\sigma^i_{\rm low}(\omega,T).
\end{equation}
Here $\sigma^i_{\rm high}(\omega,T)$ is the high energy part of the spectral functions which is
non-zero only for $\omega>2 m$ and 
describes the propagation of bound or unbound quark anti-quark pairs. On the other hand
$\sigma^i_{\rm low}(\omega,T)$ 
receives the dominant contribution
at $\omega \simeq 0$. 
Because the width of the peak at $\omega \simeq 0$ is small the later gives
an almost constant contribution to the Euclidean correlator.
We can write an analogous decomposition for the Euclidean correlator
\begin{equation}
G^i(\tau,T)=G^i_{\rm high}(\tau,T) + G^i_{\rm low}(\tau,T).
\end{equation}
To very good approximation $G^i_{\rm low}(\tau,T)=\chi^i(T) T$, i.e. constant.
In the next two sections I will discuss the temperature dependence of $G^i_{\rm high}$
and $G^i_{\rm low}$ separately.

\section{Lattice calculations of  quarkonium correlators}
Meson correlators have been calculated in quenched approximation using
standard Wilson gauge action and non-perturbatively improved Wilson action 
for heavy quarks. The analysis has been performed at three different lattices
spacings and several values of the quark masses in the region of the charm quark
mass. On the finest lattice also the case of bottomonium has been considered.
The parameters used in lattice calculations, including the lattice gauge coupling 
$\beta=6/g^2$ and hoping parameters $\kappa$ are summarized in Table \ref{tab2}.
The value of the clover constant and critical hopping parameter $\kappa_c$ are
taken from Ref. \cite{alpha} and interpolation. In the Table I also give the bare
input quark mass defined as 
\begin{equation}
a m_0=\frac{1}{2 \kappa}-\frac{1}{2 \kappa_c}
\end{equation}
At each lattice spacing meson correlators have been calculated at temperature
well below the transition temperature and for several temperature values in
the deconfined phase. The lattice volumes and the number of gauge configurations
used in the analysis below $T_c$ are also given Table \ref{tab2}.
The lattice spacing was set using the Sommer scale $r_0=0.5fm$ in contrast
to Ref. \cite{datta04} where it was fixed by the string tension.
In determining the lattice spacing the values of $r_0/a$ calculated in Ref. \cite{necco01}
were used together with a fit to Alton Ansatz for $6.1< \beta \le 6.92$. 
The lattice spacing determined this way are given in Table \ref{tab2}. 
Calculations in the deconfined phase span the temperature range between $1.1T_c$
and $3T_c$.
In determining the temperature scale $T/T_c$ I use the value $r_0 T_c=0.7498(50)$ determined
in Ref. \cite{necco04}. The lattice volumes and the number of gauge configurations
for each temperature above the transition temperatures are given in Table \ref{tab3}.

The temperature dependence of meson correlators comes from the explicit
temperature dependence of the spectral function and the trivial temperature
dependence of the integration kernel in Eq. (\ref{spectral}).
Therefore to eliminate the trivial temperature dependence of the integration
kernel we follow the procedure proposed in Ref. \cite{datta04} and calculate
the so-called reconstructed correlators
\begin{equation}
G_{rec}^i(\tau,T) = \int_0^{\infty} d \omega
\sigma^i(\omega,T^*)  \frac{\cosh(\omega(\tau-1/2 T))}{\sinh(\omega/2 T)},
\label{grec}
\end{equation}
where $T^*$ is some reference temperature well below $T_c$. Obviously if
the spectral function does not change across the deconfinement transition the ratio
$G^i/G_{rec}^i$ will be temperature independent an equal to unity.
Thus we need to calculate the spectral functions below $T_c$.
This has been done using MEM and the algorithm described in Ref. \cite{jako07}.
The position of the first peak in the spectral function gives the mass
of the lowest lying state in that channel. In Table \ref{tab2} I 
show the mass of the ground state in the pseudo-scalar channel, i.e.
the $\eta_c(1S)$ ($\eta_b(1S)$) mass. The statistical error has been
estimated using jackknife analysis of the peak position. There are also
systematic uncertainties in the MEM analysis (e.g. due to use of different
priors). Therefore in Table \ref{tab2} the combined statistical and
systematic uncertainties in the determined mass values are shown
For each input bare quark mass I estimate the constituent quark
mass by matching the $\eta_c(1S)$ ($\eta_b(1S)$) mass calculated on the lattice
to that calculated in the potential model. A Cornell type potential 
\begin{equation}
V(r)=-\frac{\pi}{12 r}+ \sigma r, 
\end{equation}
was used with $\sigma=(1.65-\pi/12)/r_0^2$ \footnote{This value of $\sigma$ follows
from the definition of the Sommer scale $r^2 V'_r(r)|_{r=r_0}=1.65$.}
This form gives a very good description
of the lattice results on the static potential in SU(3) gauge theory.
The constituent quark masses calculated this way are given in Table \ref{tab2}.
 
One also has to keep in mind that meson currents has to be renormalized if
calculated on the lattice. The renormalization constants for scalar, pseudo-scalar,
vector and axial-vector currents calculated in 1-loop tadpole improved perturbation theory
are given in Table \ref{tab2} (see Ref. \cite{datta04} for details of the calculations).
The renormalization constants in the scalar and pseudo-scalar channels depend on
the renormalization scale which was chosen to be the constituent heavy quark mass.
In the scalar channel we will also need the renormalization constant estimated
at scale equal to half and twice the constituent quark mass.

\begin{figure}
\resizebox{0.5 \textwidth}{!}{%
\includegraphics  {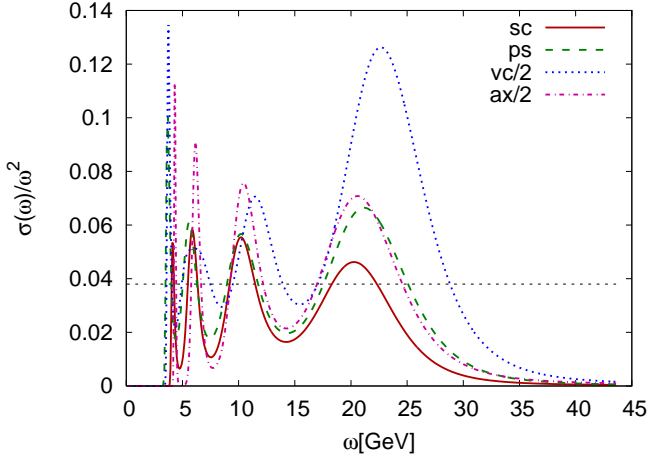}
}
\caption{The spectral functions in different channels calculated using $48^3 \times 64$ lattice
at $\beta=7.192$. The vector and axial-vector spectral functions are divided by two. The horizontal
line shows the spectral function in the massless free theory $\sigma(\omega)/\omega^2=3/( 8 \pi^2)$.}
\label{fig:spf}
\end{figure}

To demonstrate the general features of the spectral functions below $T_c$ in Fig. \ref{fig:spf}
I show the spectral functions calculated  for $\beta=7.192$. In addition to
ground state peak they show a second peak which may be due to combination of several
excited states and two broad bumps at even higher energy. The later are most likely
artifacts of limited statistics. In Ref. \cite{jako07}, where the spectral functions
have been calculated with significantly more statistics the high energy part of the
spectral functions appears to be much smoother.

As mentioned above the ratio $G(\tau,T)/G_{rec}(\tau,T)$ can be used to study
the temperature dependence of the quarkonium correlators
\footnote{In the remainder of this section the subscript $i$ labeling the channel will
be omitted. It should be clear from the context which channel is discussed.}. 
However, the temperature
dependence of the correlators can be induced by the high energy part of the 
spectral functions and the low energy part of the spectral functions, which gives
the zero mode contribution. The zero mode contribution is absent in the derivative
of the correlator with respect to $\tau$. Therefore the best way to study the 
temperature dependence of the correlators induced by change of bound state properties
and/or its dissolution is to consider the ratio of the derivatives of the correlators
$G'(\tau,T)/G_{rec}'(\tau,T)$. In the next two subsection we are going to 
discuss this ratio for pseudo-scalar channel as well as for scalar and axial-vector
channels respectively.

\subsection{The temperature dependence of the pseudo-scalar correlators}
\begin{figure}
\resizebox{0.5 \textwidth}{!}{%
\includegraphics{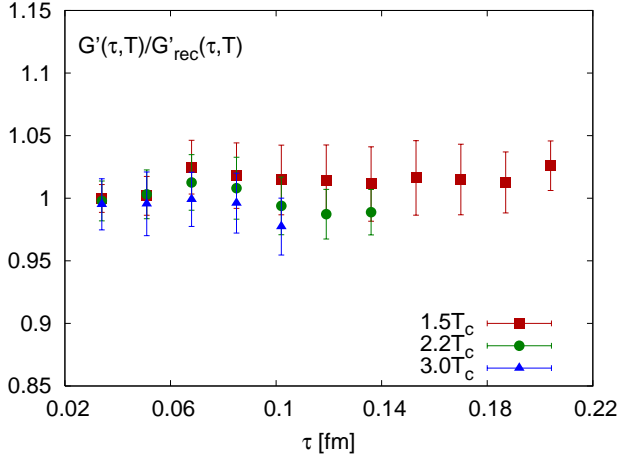}
}\\
\resizebox{0.5 \textwidth}{!}{%
\includegraphics{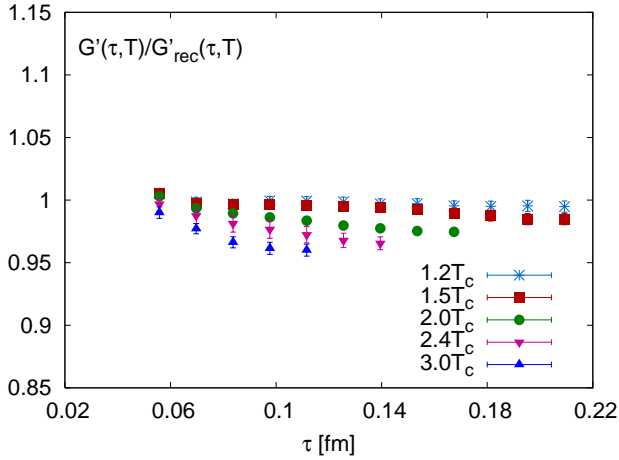}
}
\caption{The ratio of the derivatives $G'(\tau,T)/G'_{rec}(\tau,T)$ in the pseudo-scalar
channel calculated on isotropic lattices at $\beta=7.192$ (top) as well results from
anisotropic lattice calculations at $\beta=6.5$ \cite{jako07} (bottom).}
\label{fig:ratps}
\end{figure}

The temperature dependence of the charmonium and bottomonium  correlators in pseudo-scalar
channel has been studied in Refs. \cite{datta04,jako07,panic05}. In calculations on
isotropic lattices the charmonium correlators do not show any temperature dependence up
to temperatures $1.5T_c$. More precisely the ratio $G(\tau,T)/G_{rec}(\tau,T)$
is one within statistical errors up to this temperature. At higher temperatures we see
deviation from unity in this ratio. In particular at $3T_c$ this ratio deviates from one 
by about $10\%$ \cite{datta04}. 
In anisotropic lattice calculations the ratio $G(\tau,T)/G_{rec}(\tau,T)$
is also close to one up to temperatures $1.5T_c$ although statistically significant deviation from
unity are seen at the level of few percent. At higher temperatures, however, one sees significant
deviations from unity which are also larger than those observed in isotropic lattice calculations
and are present  at smaller values of $\tau$ \cite{jako07}. These deviations are larger
on coarser lattices. In the bottomonium case no temperature dependence was found up
to $3T_c$.
In Figure \ref{fig:ratps} I show the ratio of the derivatives $G'(\tau,T)/G_{rec}'(\tau,T)$
calculated at $\beta=7.192$. The results for this ratio
calculated on anisotropic lattices with larger spatial lattice spacings are also shown \cite{jako07}.
As one can see from the figure on isotropic lattices the ratio of the derivatives
does not show any temperature dependence and agrees with one within statistical errors.
The temperature dependence of the pseudo-scalar correlator seen in anisotropic lattice calculations
is also greatly reduced and the agreement between isotropic and anisotropic calculations is
is better for this ratio.
One possible explanation of this behavior is the presence of a small negative zero mode in the
pseudo-scalar channel. In the continuum there is no zero mode contribution in the pseudo-scalar
at leading order (c.f. Table \ref{tab1}) and this is true also on the lattice \cite{karsch03}.
While we do not expect to have a negative zero mode contribution even at higher orders 
in the continuum theory, the situation may be different on the lattice. 
There could be a negative zero mode contribution in the pseudo-scalar channel which vanishes
in the continuum limit
The fact that deviations
from unity in the ratio $G/G_{rec}$ increases with increasing the lattice spacing makes 
this scenario a plausible one.
Apparently modifications of bound state properties is not reflected in the temperature
dependence of the correlators, i.e. $G_{\rm high}(\tau,T)$ is temperature independent at the level
of few percent. The temperature dependence of the pseudo-scalar correlators is consistent with
potential model calculations \cite{mocsy07a,mocsy_zj75}.

\subsection{Scalar and axial-vector correlators}
\begin{figure}
\resizebox{0.5 \textwidth}{!}{%
\includegraphics{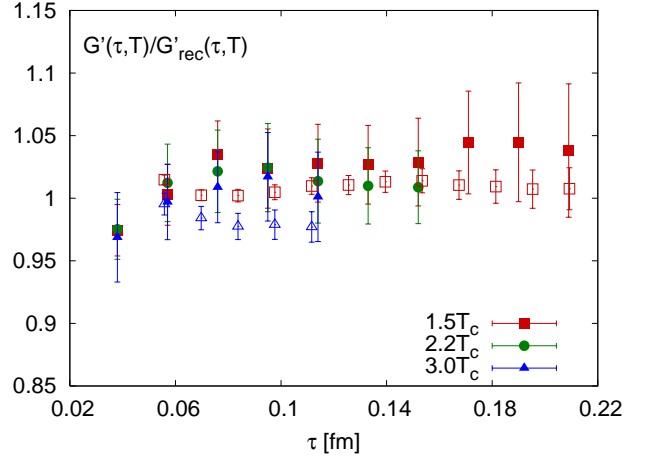}
}\\
\resizebox{0.5 \textwidth}{!}{%
\includegraphics{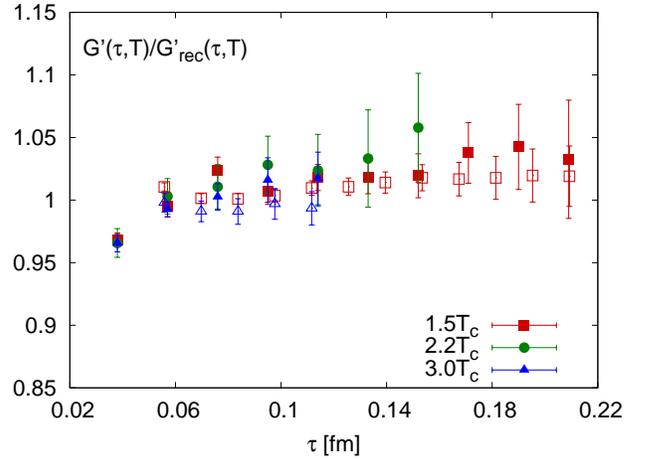}
}
\caption{The ratio of the derivatives $G'(\tau,T)/G'_{rec}(\tau,T)$ in the scalar
channel (top) and axial-vector channel (bottom) calculated at $\beta=7.192$.
The results from anisotropic lattice calculations at $\beta=6.5$ \cite{jako07}
are also shown (open symbols).}
\label{fig:ratP}
\end{figure}

The temperature dependence of scalar and axial-vector correlators has been studied
in terms of $G/G_{rec}$ using isotropic \cite{datta04} as well as anisotropic lattices \cite{jako07}.
The ratio $G/G_{rec}$ is temperature independent in the confined phase \cite{jako07} and shows
large enhancement in the deconfined phase \cite{datta04,jako07}. This large enhancement 
is present both in charmonium and bottomonium correlators. To eliminate the zero mode contribution
I have calculated the derivative of the correlators  and the corresponding ratio $G'(\tau,T)/G'_{rec}(\tau,T)$.
The numerical results for this ratio at $\beta=7.192$ are shown in Figure \ref{fig:ratP}. 
Here I also show the results from anisotropic lattices \cite{jako07}. There is good agreement between
the results obtained from isotropic and anisotropic lattices.
As one can see
from the Figure $G'(\tau,T)/G'_{rec}(\tau,T)$ shows very little temperature dependence and is close
to unity. This means that almost the entire temperature dependence of the scalar and axial-vector
correlators is due to zero mode contribution and $G_{\rm high}(\tau,T)$ is temperature independent.
One may wonder if the temperature independence of $G_{\rm high}(\tau,T)$  implies the survival
of the P-state quarkonium up to temperature as high as $3T_c$. The P-state quarkonium  has binding energy
that is significantly smaller than that of the S-state. Therefore we expect it to melt at lower
temperatures. Indeed, potential model calculations show that the dissociation temperatures for the P-states
is significant lower than for S-states and none of the potential models can 
get dissociation temperatures as high as $3T_c$ for P-states. 
The potential model calculations 
\cite{karsch88,ropke88,hashimoto88,digal01a,digal01b,digal02,shuryak04,wong04,blaschke,alberico,mocsy06,rapp,alberico06,mocsy07a,mocsy07b,mocsy_sqm07} 
take into account the temperature dependence of the real part of the
potential. At finite temperature the potential has also an imaginary part \cite{laine1,brambilla08}.
The imaginary part of the potential further weakens the bound state contribution \cite{laine2}.
The very weak temperature dependence of the quarkonium correlators despite of melting of the bound states
looks puzzling at a first glance.
However, the calculation of quarkonium
correlators within potential model shows that the high energy part of the correlator almost
does not change with the temperature even when the P-state quarkonium melts \cite{mocsy07a}.

Note that the temperature independence of meson correlators is true only for heavy quarks. In
the light quark sector meson correlators show significant temperature dependence \cite{karsch02,me_sqm03}.

\section{Estimating the zero mode contribution}
\begin{figure}
\resizebox{0.5 \textwidth}{!}{%
\includegraphics{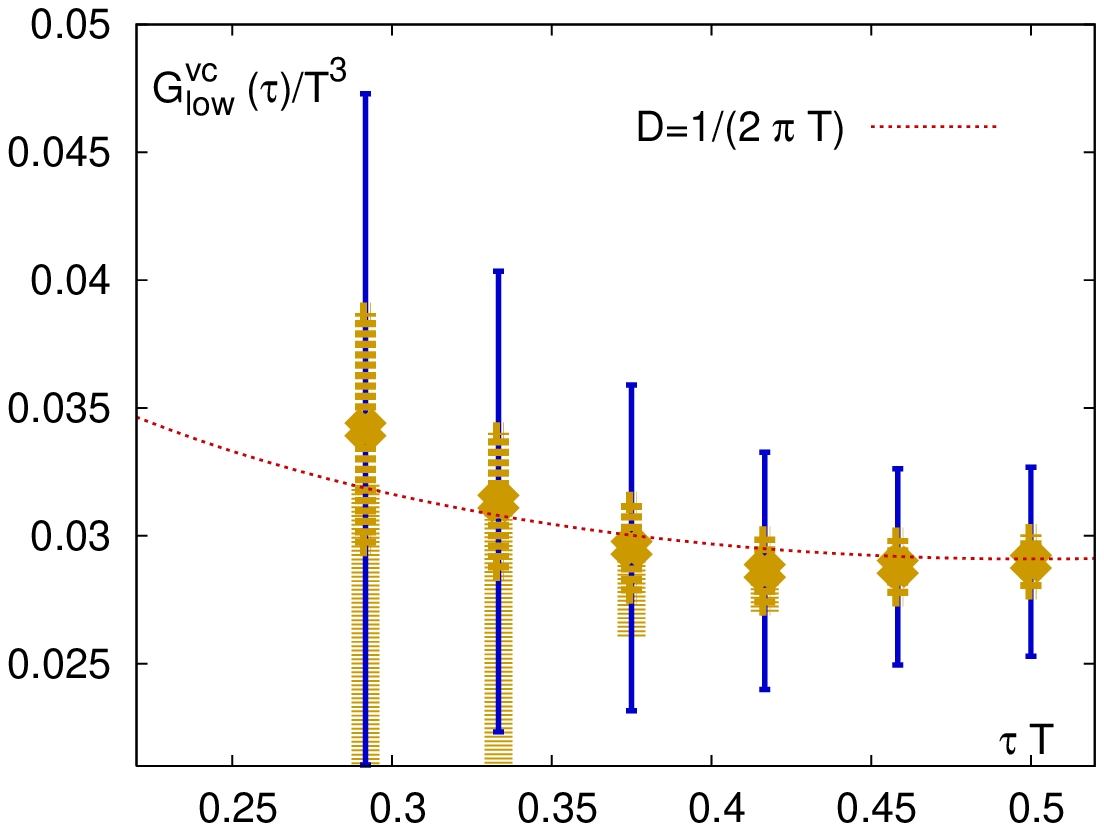}}\\
\resizebox{0.5 \textwidth}{!}{%
\includegraphics{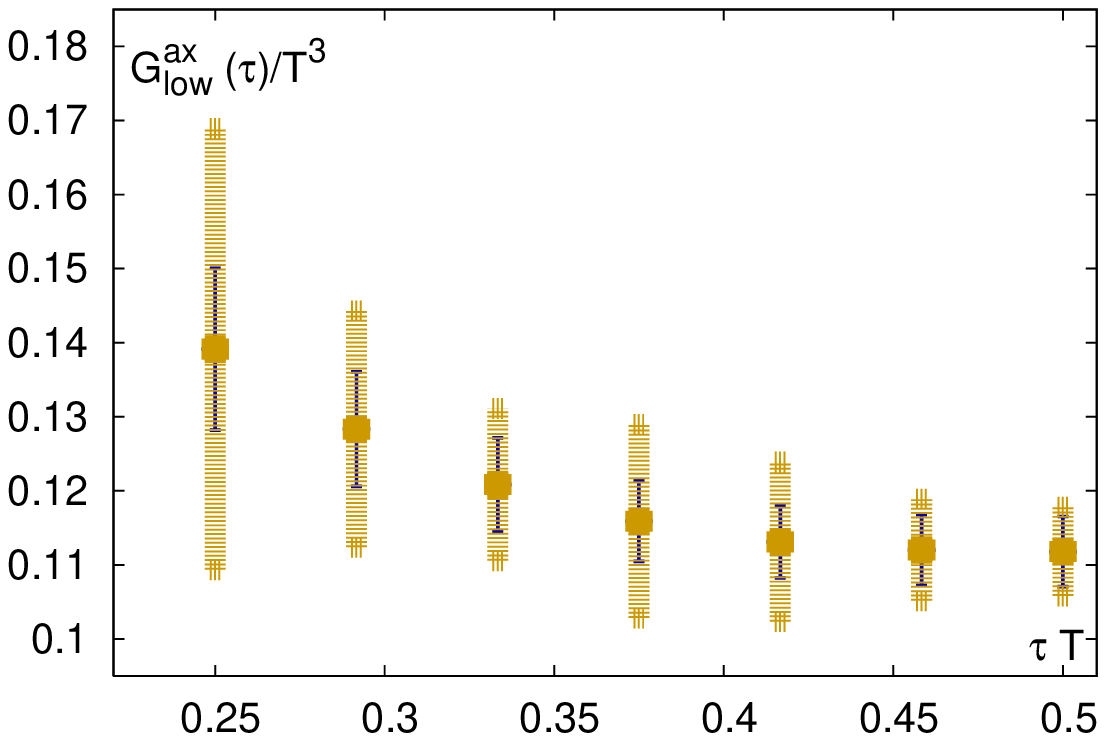}
}
\caption{
The low energy part of the vector (top) and
axial-vector (bottom) correlators as function
of $\tau$. The thin error bars correspond to statistical
error, while the thick error bars to systematic errors
(see the main text).
}
\label{fig:zeromode}
\end{figure}

\begin{figure}
\resizebox{0.5 \textwidth}{!}{%
\includegraphics{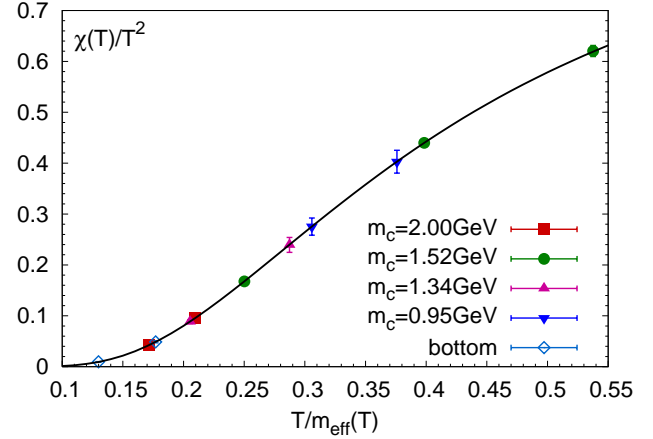}}\\
\resizebox{0.5 \textwidth}{!}{%
\includegraphics{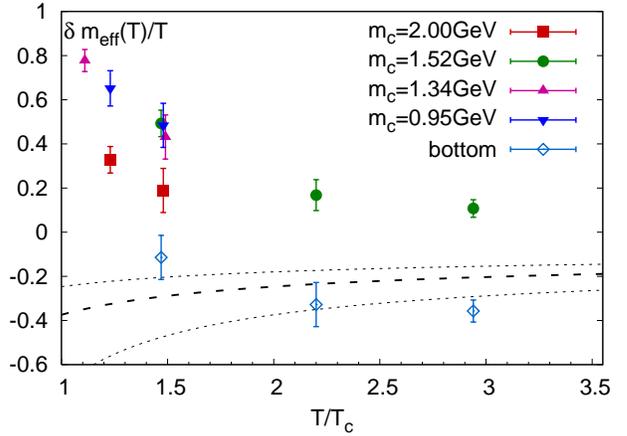}
}
\caption{
The quark number susceptibility as function
of  $T/m_{\rm eff}(T)$ (top) and the thermal mass
correction (bottom) as function of the temperature.
The solid black line corresponds to the quasi-particle
model. The dashed line and the band correspond to the
perturbative prediction of the thermal mass correction.
}
\label{fig:chi}
\end{figure}

\begin{table}
\caption{Finite temperature lattices and the corresponding
number of gauge configurations used in the analysis
of quarkonium correlators in the deconfined phase.
For each lattice volume the corresponding value of the
temperature is also shown.}
\begin{tabular}{llll}
\hline
$\beta$ & lattice             & \#config. & $T/T_c$ \\
6.499   & $48^3 \times 12$    &   50       & 1.23   \\
        & $48^3 \times 10$    &   46       & 1.48   \\
\hline
6.640  &  $48^3 \times 16$    &   50       & 1.11   \\
       &  $48^3 \times 12$    &   60       & 1.49   \\
\hline 
7.192  &  $64^3 \times 24$    &   80       & 1.47   \\
       &  $48^3 \times 16$    &   100      & 2.20   \\
       &  $48^3 \times 12$    &   90       & 2.94   \\
\hline  
\end{tabular}
\label{tab3}
\end{table}

\begin{figure}
\resizebox{0.5 \textwidth}{!}{%
\includegraphics{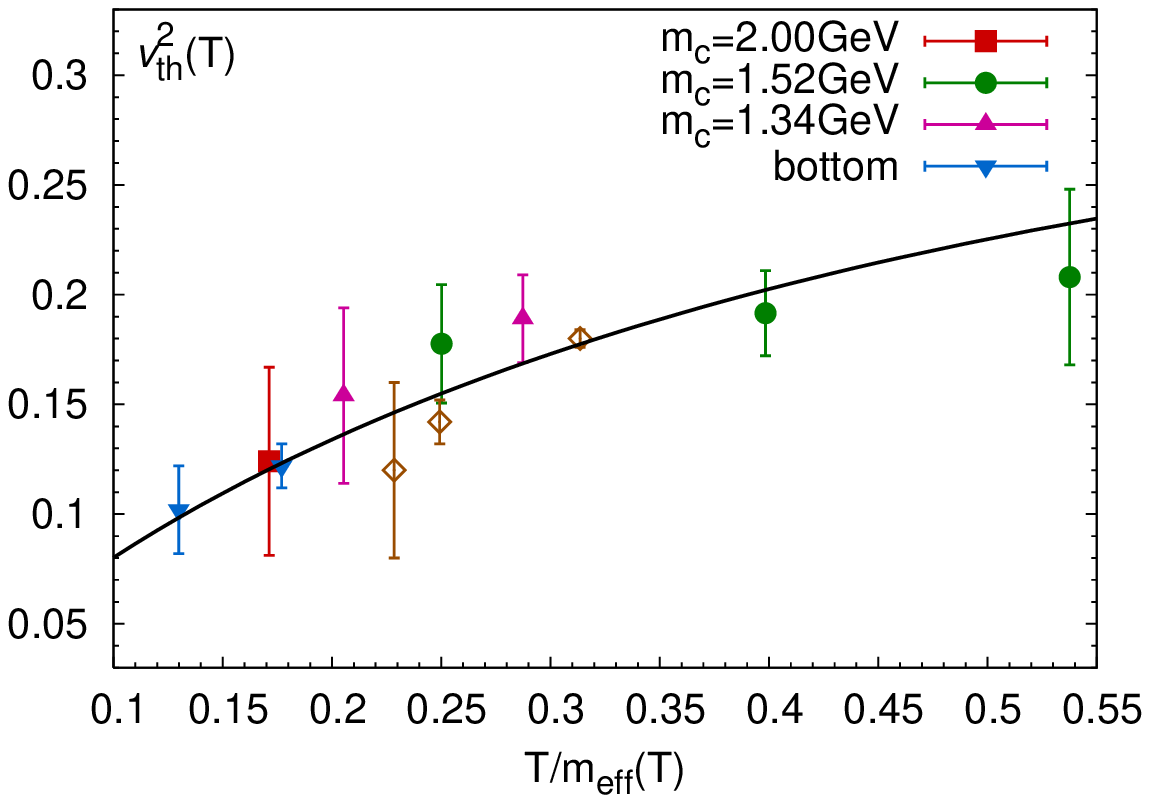}}\\
\resizebox{0.5 \textwidth}{!}{%
\includegraphics{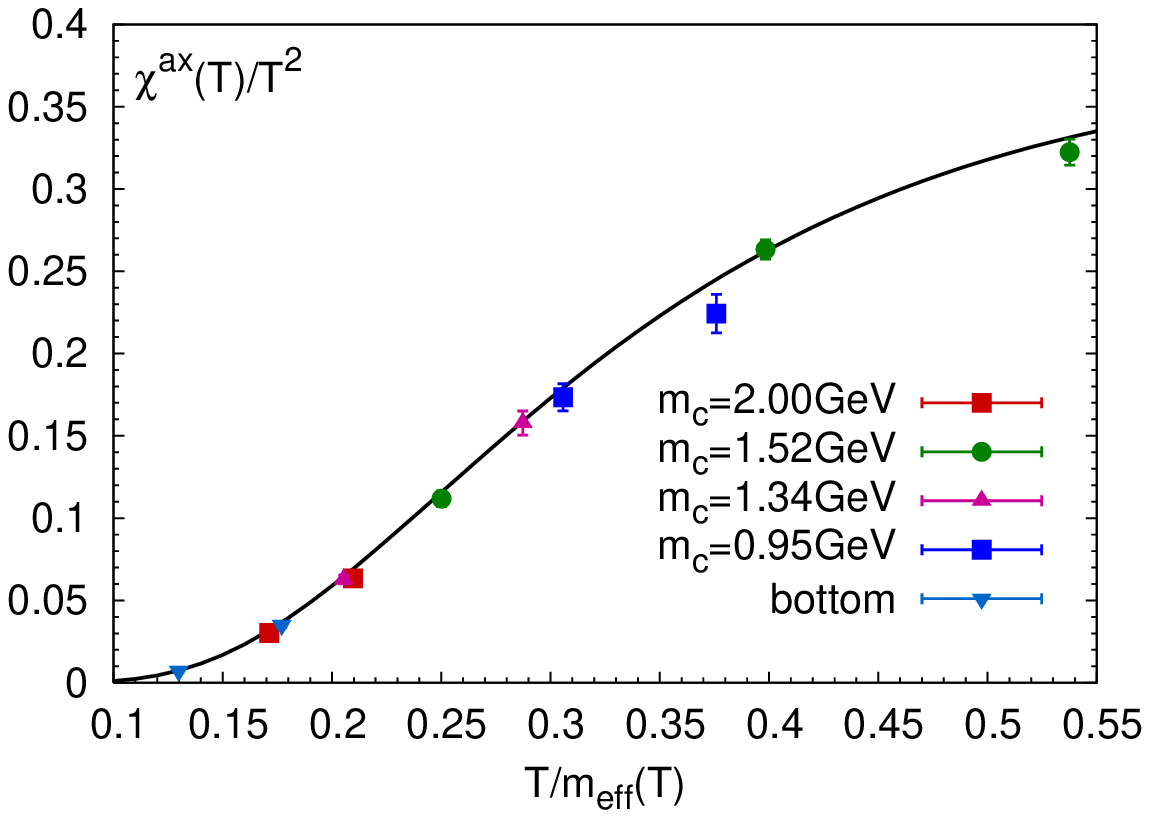}
}
\caption{
The thermal velocity of heavy quarks (top) and the zero mode 
contribution to the axial-vector correlator $G^{ax}_{\rm low}/T^3=\chi^{ax}/T^2$ (bottom). The lines show
the prediction of the quasi-particle model with $m_{\rm eff}(T)$.
The open symbols show the thermal velocity squared estimated on anisotropic lattices \cite{jako07}.
}
\label{fig:zeroaxvc}
\end{figure}

As discussed in the previous section the high energy part of the quarkonium
correlators turns out to be temperature independent to very good approximation.
Therefore we could use the spectral functions calculated below $T_c$ to estimate it, i.e.
we can assume that  $G^i_{\rm high}(\tau,T) \simeq G^i_{rec}(\tau, T)$.
Then the low energy part of the correlators, i.e. the zero mode contribution can
be evaluated as $G^i_{\rm low}(\tau,T) = G^i(\tau,T)-G^i_{rec}(\tau,T)$.
It is important to estimate the systematic errors in this difference due to inaccuracies
in the determination of the spectral functions 
in the confined phase, i.e. $G_{rec}^i(\tau,T)$, as well as due to the small temperature dependence of
$\sigma_{\rm high}(\omega)$. To estimate the errors in the reconstructed correlator
I compared several spectral functions calculated using different Bayesian priors and different
jackknife sub-samples. The uncertainties in $G_{rec}^i(\tau,T)$ turn out to be smaller than $1\%$.
To estimate the uncertainties due to 
temperature dependence of $\sigma_{\rm high}(\omega)$, i.e. due to the fact that it is slightly different
from the zero temperature spectral functions we consider the pseudo-scalar channel.
Here we do not expect to have a zero mode contribution in the continuum limit and 
therefore $G^i(\tau,T)-G^i_{rec}(\tau,T) \simeq 0$. At $1.5T_c$ this difference is indeed zero within
statistical errors and we used it to estimate the systematic errors in other channels.
At higher temperatures as discussed in the previous section we may have a negative zero
mode contribution in the pseudo-scalar channel at finite lattice spacing. But this zero mode contribution
is much smaller than the zero mode contribution in other channels.
 
In Figure \ref{fig:zeromode} I show the estimated zero mode contribution $G^i_{\rm low}(\tau,T)/T^3$ for
the vector and axial-vector channels.
In the figure both statistical (thin error bars) and systematic (thick error bars) uncertainties are shown.
Both uncertainties increase with decreasing $\tau$. This is due to the fact that the relative 
contribution of the zero mode decreases fast at small Euclidean times.

To the first approximation $G^i_{\rm low}(\tau,T)/T^3$ should be
constant. 
We see that the numerical results for this quantity are indeed compatible with constant behavior
in $\tau$ within the estimated errors. The curvature of $G^i_{\rm low}(\tau,T)$ is controlled by the diffusion
constant $D$ \cite{derek05}. However, even for the smallest possible value of the diffusion constant $D=1/(2 \pi T)$ (one
thermal wavelength) the curvature of  $G^i_{\rm low}(\tau,T)$ is quite small as can be seen from
the Figure \ref{fig:zeromode}. Current lattice data are clearly not precise enough to constrain the value of $D$, they
can only tell about the area under the peak at $\omega \simeq 0$.
Therefore we could take the value of low energy part of the correlator at the midpoint as an estimate for
the zero mode contribution, $G_{\rm low}^i(\tau=1/(2T),T)=T \chi^i(T)$. In the following I will discuss the temperature
dependence of $\chi^i(T)$ for different quark masses and in different channels. 

Since the zero mode contribution is related to the propagation
of single (unbound) heavy quark in the medium it is natural to ask to what extent its temperature dependence  can be
described in terms of a  quasi-particle model.  For this purpose
let us consider the temporal component of the vector correlator $G^{vc0}$. In this
case there is no high energy component
and $G^{vc0}= - T \chi^{vc0}(T) \equiv - T \chi(T)$, with $\chi(T)$ being the heavy quark number susceptibility. 
In the free theory
$\chi/T^2$ depends only on $m/T$.
Thus matching the free theory expression for $\chi^{vc0}\equiv \chi(T)$ (Eq. (\ref{gen_susc})) 
to the lattice data on $G^{vc0}$, we can define
and effective temperature dependent heavy quark mass $m_{\rm eff}(T)$.
It is also convenient to define an effective thermal mass correction
$\delta m_{\rm eff}(T)=m_{\rm eff}(T)-m$.
In Figure \ref{fig:chi} I show the quark number susceptibility as function of $T/m_{\rm eff}$ as well
as the effective thermal mass correction $\delta m_{\rm eff}(T)$.
The thermal correction to the heavy quark mass is largest at $T_c$ and monotonically decreases with 
increasing temperature. It also depends on the heavy quark mass, namely it decreases with increasing quark mass.
For very heavy quarks the thermal mass correction is given by \cite{laine1,brambilla08,beraudo}
\begin{equation}
\delta m_{\rm eff}=-\frac{4}{3} \frac{g^2(T)}{4 \pi} m_D,
\end{equation} 
with $m_D=g(T) T$ being the perturbative Debye mass ($N_f=0$ because we work in quenched approximation).
It is interesting to compare this
leading order perturbative prediction with the numerical results on $\delta m_{\rm eff}(T)$. For this
we have to determine the running coupling $g(T)$. I used the 2-loop running coupling evaluated
at scale $\mu=2 \pi T$ as well as the relation $T_c/\Lambda_{\overline{MS}}=1.15(5)$ \cite{sorendu}.
The corresponding perturbative predictions are shown as the dashed line in Figure \ref{fig:chi}.
To estimate the uncertainties in this perturbative prediction I also evaluated the coupling at
$\mu=\pi T$ and $\mu=4 \pi T$. This is shown as the band in Figure \ref{fig:chi}.

Now I discuss the temperature dependence of the zero mode contribution in the vector and axial-vector
channels. In general the zero mode contribution depends on both the temperature and the quark mass.
In quasi-particle picture, however, it is a function of $T/m_{\rm eff}$ only. Therefore in 
Figure \ref{fig:zeroaxvc} I show the numerical results for $\chi^{ax}(T)$ as function of $T/m_{\rm eff}$.
Here I used the temperature dependent effective quark masses determined above. We see that $\chi^{ax}(T)/T^2$
is indeed function of $T/m_{eff}$ only and agrees very well with the prediction of the quasi-particle model
shown in the Figure as the black line.

Similar
analysis has been done in the vector channel. In this case I analyzed the ratio 
$G_{\rm low}^{vc}(T)/G^{vc0}$ which has a simple physical interpretation as the averaged thermal velocity squared
of the heavy quark $v_{\rm th}^2(T)$.
This is because due to the heavy quark mass the Fermi-Dirac distribution in Eq. (\ref{gen_susc}) can be replaced
by the Boltzmann distribution and we have  
\begin{equation}
\frac{G_{\rm low}^{vc}(T)}{G^{vc0}(T)} \simeq 
\left(\int d^3p \frac{p^2}{E_p^2} e^{-E_p/T}\right)/\left( \int d^3 p e^{-E_p/T} \right)=v_{\rm th}^2.
\end{equation}
The numerical results for $v_{th}^2(T)\equiv G_{\rm low}^{vc}(T)/G^{vc0}(T)$ are shown in Figure \ref{fig:zeroaxvc}.
As one can see also  $v_{th}^2(T)$ is a function of $T/m_{\rm eff}$ only and agrees with the prediction of
a quasi-particle model shown as the black curve. The zero mode contribution in the vector channel has also
been studied on anisotropic lattices \cite{jako07} and in Figure \ref{fig:zeroaxvc} I also show the corresponding
results for thermal velocity squared. To estimate $m_{\rm eff}$ in this case I used $1.19GeV$ for the constituent
charm quark mass needed to get the charmonium spectrum right in the potential model \cite{mocsy07a} and 
$\delta m_{\rm eff}(T)$ determined above.   
For $T \gg m$ we have $v_{\rm th}^2 \simeq T/m$ but as one can see from the figure
it is not a good approximation even for the b-quark.

\begin{figure}
\resizebox{0.5 \textwidth}{!}{%
\includegraphics{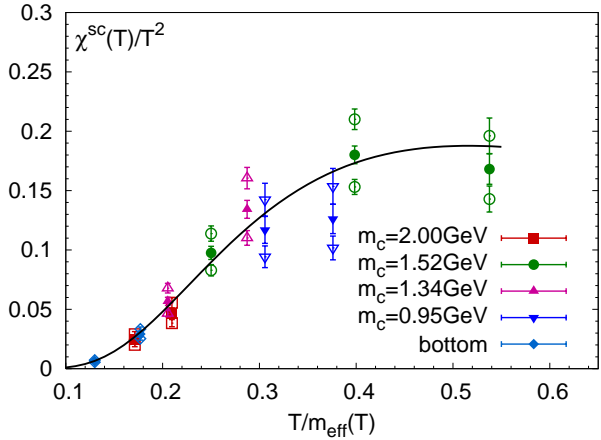}}
\caption{
The zero mode contribution to the scalar correlator in units of $T^3$
as function of $T/m_{\rm eff}(T)$. The black line is the prediction
of quasi-particle model. The filled symbols correspond to the scalar
correlator evaluated at scale $\mu=m$, while the open symbols 
correspond to the scalar correlator evaluated at scale $\mu=m/2$ and
$\mu=2m$. 
}
\label{fig:zerosc}
\end{figure}

Finally I discuss the zero mode contribution in the scalar channel. Unlike the vector and axial-vector correlators
the scalar correlator depends on the choice of the renormalization scale. It is not obvious what is the suitable choice of
the scale when comparing the zero mode contribution to the prediction of the quasi-particle model.
Since the zero mode contribution is related to the propagation of a single quark the choice $\mu=m$ for the renormalization
scale seems to be a natural one. To get an idea about the sensitivity of the conclusions to the choice of the
renormalization scale I have also considered $\mu=m/2$ and $\mu=2 m$. Using the renormalization constants given
in Table \ref{tab2}. the zero mode contribution in the scalar channel has been evaluated at scales $\mu=m/2,~m,~2m$.
The numerical results for the zero mode contribution to the scalar correlator as function of $T/m_{\rm eff}$
are shown in Figure \ref{fig:zerosc}. We see again that the zero mode contribution is function of
$T/m_{\rm eff}$ only and there is a reasonable agreement with the prediction of the quasi-particle model shown
as a black line.

\begin{figure}
\resizebox{0.5 \textwidth}{!}{%
\includegraphics{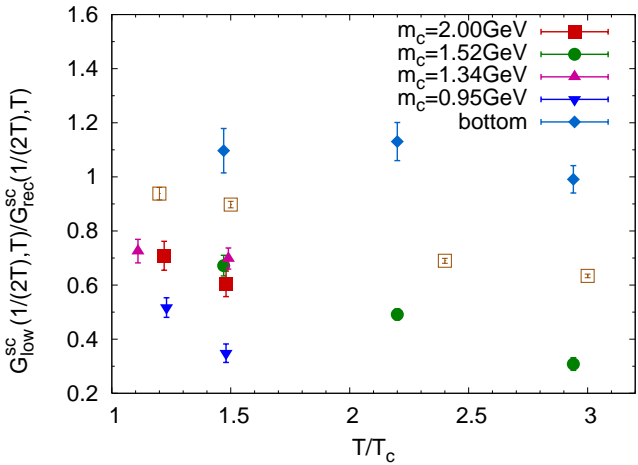}}
\caption{The ratio of the zero mode contribution of the scalar correlator
to the corresponding reconstructed correlator at $\tau=1/(2T)$.
Open symbols correspond to the results obtained on anisotropic lattices \cite{jako07}.
}
\label{fig:zerosc1}
\end{figure}

One way to quantify the zero mode contribution to the scalar correlator which is independent
of the choice of the renormalization scale is to consider the ratio $G_{\rm low}^{sc}/G_{rec}^{sc}$
evaluated at $\tau=1/(2 T)$. The numerical results for this ratio are shown in Figure \ref{fig:zerosc1}.
As one can see from the figure this quantity decreases with increasing temperature and decreasing quark masses.
There is no large dependence lattice spacing  in this quantity.
In the Figure I also show the results obtained on anisotropic lattices at $\beta=6.5$ \cite{jako07},
which are significantly larger than the results obtained on isotropic lattice. This is most likely
due to the fact that the spatial lattice spacing at $\beta=6.5$ is too coarse for studying the
zero mode contribution.

\section{Conclusions}
In this paper the temperature dependence of quarkonium correlation function has been discussed. It has been shown
that the temperature dependence of the high energy part of the spectral function, for example, the melting of 
resonances does
not lead to a large change in the correlation function. 
The dominant source of the temperature dependence of quarkonium
correlators is the zero mode contribution. This contribution has been studied quantitatively on the lattice 
for the first time. In general, it  is expected that this contribution  depends on the temperature and the quark mass. 
We have found, however,  that it is the function
of $m_{\rm eff}/T$ only and is well described by a quasi-particle model down to temperatures as low as $1.1T_c$.
Since the the quasi-particle mode is so successful in describing the zero mode contribution to
the quarkonium correlators it would be 
of great interest to calculate it systematically in improved perturbation theory \cite{scpt}.

\section*{Acknowledgments}
This work was supported by U.S. Department of Energy under
Contract No. DE-AC02-98CH10886. I am grateful to A. Jakov\'ac for providing his code
for the Maximum Entropy Method analysis and S. Datta for participation at early stages
of this work.


\begin{thebibliography}{}
\bibitem{MS86}
T.~Matsui and H.~Satz,
Phys.\ Lett.\ B {\bf 178} (1986) 416
\bibitem{karsch88}
F.~Karsch, M.~T.~Mehr and H.~Satz,
Z.\ Phys.\ C {\bf 37} (1988) 617
 
\bibitem{ropke88}
G. R{\"o}pke, D. Blaschke, H. Schulz, Phys.\ Rev.\ D {\bf 38} (1988) 3589 
 
\bibitem{hashimoto88}
T.  Hashimoto et al., Z.\ Phys.\ C {\bf 38} (1988) 251 
 
\bibitem{digal01a}
S.~Digal, P.~Petreczky and H.~Satz,
Phys.\ Lett.\ B {\bf 514} (2001) 57 
[arXiv:hep-ph/0105234].

\bibitem{digal01b}
S.~Digal, P.~Petreczky and H.~Satz,
Phys.\ Rev.\ D {\bf 64} (2001) 094015 
[arXiv:hep-ph/0106017].
 
\bibitem{digal02} 
  P.~Petreczky,
  Pramana {\bf 60} (2003) 829
  [arXiv:hep-ph/0201096].
                                                                                                
\bibitem{shuryak04}
E.~V.~Shuryak and I.~Zahed,
Phys.\ Rev.\ D {\bf 70} (2004) 054507 
[arXiv:hep-ph/0403127].
                                                                                                  
\bibitem{wong04}
  C.~Y.~Wong,
  Phys.\ Rev.\ C {\bf 72} (2005) 034906 
[arXiv:hep-ph/0408020].

\bibitem{blaschke}
  D.~Blaschke, O.~Kaczmarek, E.~Laermann and V.~Yudichev,
  Eur.\ Phys.\ J.\  C {\bf 43} (2005) 81
  [arXiv:hep-ph/0505053].


\bibitem{alberico}
  W.~M.~Alberico, A.~Beraudo, A.~De Pace and A.~Molinari,
  Phys.\ Rev.\ D {\bf 72} (2005) 114011 
[arXiv:hep-ph/0507084].

\bibitem{mocsy06}
  \'A.~M\'ocsy and P.~Petreczky,
  Phys.\ Rev.\  D {\bf 73} (2006) 074007
  [arXiv:hep-ph/0512156].


\bibitem{rapp}
  D.~Cabrera and R.~Rapp,
  Phys.\ Rev.\  D {\bf 76}, 114506 (2007)
  [arXiv:hep-ph/0611134].
                                                                                                  
\bibitem{alberico06}
  W.~M.~Alberico, A.~Beraudo, A.~De Pace and A.~Molinari,
  Phys.\ Rev.\  D {\bf 75} (2007) 074009 
  [arXiv:hep-ph/0612062].


\bibitem{mocsy07a}
  \'A.~M\'ocsy and P.~Petreczky,
  Phys.\ Rev.\  D {\bf 77} (2008) 014501
  [arXiv:0705.2559 [hep-ph]].

\bibitem{mocsy07b}
  \'A.~M\'ocsy and P.~Petreczky,
  Phys.\ Rev.\ Lett.\  {\bf 99} (2007) 211602
  [arXiv:0706.2183 [hep-ph]].

\bibitem{mocsy_sqm07}
  \'A.~M\'ocsy and P.~Petreczky,
  J.\ Phys.\ G {\bf 35} (2008) 044038.

\bibitem{mocsy_zj75}
  A.~Mocsy and P.~Petreczky,
  Eur.\ Phys.\ J.\ ST {\bf 155} (2008) 101
  [arXiv:0710.5125 [hep-ph]].

\bibitem{me_hard04}
  P.~Petreczky,
  Eur.\ Phys.\ J.\  C {\bf 43} (2005) 51
  [arXiv:hep-lat/0502008].


\bibitem{olaf_cpod07}
  O.~Kaczmarek,
  PoS  {\bf CPOD07} (2007) 043
  [arXiv:0710.0498 [hep-lat]].


\bibitem{umeda02}
T.~Umeda, K.~Nomura and H.~Matsufuru,
Eur.\ Phys.\ J.  C {\bf 39S1} (2005) 9 
[arXiv:hep-lat/0211003].
                                                                                                  
\bibitem{asakawa04}
M.~Asakawa and T.~Hatsuda,
Phys.\ Rev.\ Lett.\  {\bf 92} (2004) 012001 
[arXiv:hep-lat/0308034].
                                                                                                  
\bibitem{datta04}                                                                                     
S.~Datta, F.~Karsch, P.~Petreczky and I.~Wetzorke,
Phys.\ Rev.\ D {\bf 69} (2004) 094507 
[arXiv:hep-lat/0312037].

\bibitem{doi}   
  H.~Iida, T.~Doi, N.~Ishii, H.~Suganuma and K.~Tsumura,
  Phys.\ Rev.\  D {\bf 74} (2006) 074502 
  [arXiv:hep-lat/0602008].


\bibitem{me_sqm06}
  P.~Petreczky,
  J.\ Phys.\ G {\bf 32} (2006) S293
  [arXiv:hep-lat/0606007].


\bibitem{jako07}
  A.~Jakov\'ac, P.~Petreczky, K.~Petrov and A.~Velytsky,
  Phys.\ Rev.\  D {\bf 75} (2007) 014506
  [arXiv:hep-lat/0611017].


\bibitem{swan}
  G.~Aarts, C.~Allton, M.~B.~Oktay, M.~Peardon and J.~I.~Skullerud,
  Phys.\ Rev.\  D {\bf 76} (2007) 094513
  [arXiv:0705.2198 [hep-lat]].


\bibitem{derek05}
  P.~Petreczky and D.~Teaney,
  Phys.\ Rev.\  D {\bf 73} (2006) 014508
  [arXiv:hep-ph/0507318].


\bibitem{umeda07}
  T.~Umeda,
  Phys.\ Rev.\  D {\bf 75} (2007) 094502
  [arXiv:hep-lat/0701005].

\bibitem{panic05}
  S.~Datta, A.~Jakov\'ac, F.~Karsch and P.~Petreczky,
  AIP Conf.\ Proc.\  {\bf 842}, 35 (2006)
  [arXiv:hep-lat/0603002].

\bibitem{me_sqm07}
  P.~Petreczky,
  J.\ Phys.\ G {\bf 35} (2008) 044033
  [arXiv:0710.5561 [nucl-th]].

\bibitem{me_qm08}
  S.~Datta and P.~Petreczky,
J. Phys. G {\bf 35} (2008) 104114
  [ arXiv:0805.1174 [hep-lat]]

\bibitem{aarts05}
  G.~Aarts and J.~M.~Martinez Resco,
  Nucl.\ Phys.\  B {\bf 726} (2005) 93
  [arXiv:hep-lat/0507004].

\bibitem{alpha}
  M.~Luscher, S.~Sint, R.~Sommer and H.~Wittig,
  Nucl.\ Phys.\  B {\bf 491} (1997) 344
  [arXiv:hep-lat/9611015].



\bibitem{necco01}
  S.~Necco and R.~Sommer,
  Nucl.\ Phys.\  B {\bf 622} (2002) 328
  [arXiv:hep-lat/0108008].



\bibitem{necco04}
  S.~Necco,
  Nucl.\ Phys.\  B {\bf 683} (2004) 137
  [arXiv:hep-lat/0309017].


\bibitem{karsch03}
  F.~Karsch, E.~Laermann, P.~Petreczky and S.~Stickan,
  Phys.\ Rev.\  D {\bf 68} (2003) 014504
  [arXiv:hep-lat/0303017].

\bibitem{laine1}
  M.~Laine, O.~Philipsen, P.~Romatschke and M.~Tassler,
  JHEP {\bf 0703} (2007) 054
  [arXiv:hep-ph/0611300];
  M.~Laine, O.~Philipsen and M.~Tassler,
  JHEP {\bf 0709} (2007) 066
  [arXiv:0707.2458 [hep-lat]].


\bibitem{brambilla08}
  N.~Brambilla, J.~Ghiglieri, A.~Vairo and P.~Petreczky,
  Phys.\ Rev.\  D {\bf 78} (2008) 014017
  [arXiv:0804.0993 [hep-ph]].

\bibitem{laine2}
\bibitem{Laine:2007gj}
  M.~Laine,
  JHEP {\bf 0705} (2007) 028
  [arXiv:0704.1720 [hep-ph]];
  Y.~Burnier, M.~Laine and M.~Vepsalainen,
  JHEP {\bf 0801} (2008) 043
  [arXiv:0711.1743 [hep-ph]]
  


\bibitem{karsch02}
  F.~Karsch, E.~Laermann, P.~Petreczky, S.~Stickan and I.~Wetzorke,
  Phys.\ Lett.\  B {\bf 530} (2002) 147
  [arXiv:hep-lat/0110208].

\bibitem{me_sqm03}
  P.~Petreczky,
  J.\ Phys.\ G {\bf 30} (2004) S431
  [arXiv:hep-ph/0305189]

\bibitem{beraudo}
  A.~Beraudo, J.~P.~Blaizot and C.~Ratti,
  Nucl.\ Phys.\  A {\bf 806} (2008) 312
  [arXiv:0712.4394 [nucl-th]]

\bibitem{sorendu}
S. Gupta, Phys. Rev. D {\bf 64} (2001) 034507

\bibitem{scpt}
  F.~Karsch, et al,
  Phys.\ Lett.\  B {\bf 401} (1997) 69;
  J.~O.~Andersen et al,
  Phys.\ Rev.\ Lett.\  {\bf 83} (1999) 2139;
  J.~P.~Blaizot et al,
  Phys.\ Rev.\ Lett.\  {\bf 83} (1999) 2906

\end{thebibliography}
\end{document}